\documentstyle[aps,epsfig]{revtex}

\begin{document}
\draft
\def\be{\begin{equation}}
\def\ee{\end{equation}} 
\def\bfi{\begin{figure}}
\def\efi{\end{figure}}
\def\bea{\begin{eqnarray}}
\def\eea{\end{eqnarray}}
\title{Computer simulations of domain growth in off-critical quenches
of two-dimensional binary mixtures}

\author{Antonio Lamura$^{1,2}$, Giuseppe Gonnella$^1$}

\address{$^1$Istituto Nazionale per la  Fisica della Materia, Unit\`a di Bari
{\rm and} Dipartimento di Fisica, Universit\`a di Bari, {\rm and}
Istituto Nazionale di Fisica Nucleare, Sezione di Bari, via Amendola
173, 70126 Bari, Italy\\
$^2$Istituto Applicazioni Calcolo, CNR, Sezione di Bari, Via Amendola 122/I,
70126 Bari, Italy \footnote{Present address}}

\maketitle
 
\begin{abstract}
The phase separation of two-dimensional binary mixtures has been studied
through numerical Langevin simulations based on a Ginzburg-Landau free
energy. 
We have considered not symmetric mixtures with and without imposed shear flow. 
In the sheared 
case our main results are as follows: (1) domains are distorted by the flow; 
(2) the structure factor has four peaks; (3) excess 
viscosity shows a peak whose position is independent of shear rate but its
height decreases increasing shear rate.  
\end{abstract}
 

\section{Introduction} \label{intro}

Fluid systems can be described at different scales.
Molecular dynamics simulations have the advantage of taking into 
account elementary forces and microscopic details providing a 
realistic description of the fluid.
On the other hand, coarse grained models, where for example a coarse-grained
variable describes the concentration of the fluid over a  mesoscopic region,
are more convenient from a computational point of view, allowing simulations
on time scales sometimes unreachable with molecular dynamics.

In phase separation of binary mixtures, for example, after the system is 
quenched by an initially disordered state into the coexistence region, 
the typical size of the  domains  of the different fluid components 
grows with power laws  which are  generally better measured
 in continuum coarse grained models\cite{bray2}.
Due to this reason,  in this paper we will
 study the phase separation of binary mixtures subjected to a shear 
flow describing the system by a continuous field that  represents 
the difference in concentration between the two phases.
Assuming local equilibrium, a chemical potential derived by a 
free-energy functional of the concentration field will be the source
of the dynamical evolution. (This model is sometimes referred in
literature as model B\cite{halperin}). 
A convection term due the presence of the shear flow will complete the
description of the system.

The problem of phase separation of binary mixtures in a shear flow
is relevant for many practical reasons especially concerning 
polymer mixtures and dilute polymer solutions.
New recent results have shown that the shear flow greatly affects the 
morphology  of the growing 
domains\cite{ohta,rothman,olson,wu,padilla,shou}. 
Not only  the growth is anisotropic
with interfaces aligning with the direction of the flow, but for each 
direction two typical lenghtscales have been measured\cite{corberi1,corberi2}. 
This last feature
makes a deep difference with what occurs in phase separation
without imposed flows where the whole domain evolution is characterized
by a single lengthscale $R(t)$\cite{bray2}. 

In this paper we consider the case
not studied before of mixtures with off-symmetric composition.
Our model will be properly defined in Section II. 
The results will be described in Section III. Domains with two
lenghtscales for each direction will be found also for the off-symmetrical
mixtures. The role of the flow in the first stage of the phase separation
will be examined together with the behaviour of the excess viscosity.
Some conclusions will end the paper.

\section{The model} \label{model}

In the following we model a binary mixture
by coupling a
diffusive  field $\varphi$, representing  the normalized concentration
difference between the two components of the mixture, to an applied
velocity field.
This approach neglects hydrodynamic effects.
For weakly sheared polymer blends with large polymerization index 
and similar mechanical 
properties of the two species
the present model is expected 
to be satisfactory in a preasymptotic time domain  when
velocity fluctuations are small\cite{bray2}.
When hydrodynamic effects become important, instead, the full dynamical 
model\cite{wagner,cates,lamura}, where the time evolution equation for
the field $\varphi$ is
coupled to the Navier-Stokes equation, must be considered.

The mixture is described by the Langevin equation 
\begin{equation}
\frac {\partial \varphi} {\partial t} + \vec \nabla \cdot (\varphi \vec v) =
\Gamma \nabla^2  \frac {\delta {\cal F}}{\delta \varphi} + \eta
\label{eqn2}
\end{equation}
where $\Gamma$ is a mobility coefficient,
$\vec v$ is the external velocity field describing plane shear flow  
with  profile $\vec v = \gamma y \vec e_x$\cite{onuki},
$\gamma$ and $\vec e_x$ being, respectively, the  shear rate 
and the  unit vector in the $x$ direction (flow direction).
$\eta$ is a gaussian white noise, representing the effects of thermal 
fluctuations with
zero mean and correlations that, according to the fluctuation-dissipation 
theorem,
are given by
\begin{equation}
\langle \eta(\vec r, t) \eta(\vec r', t')\rangle = 
                               -2 T \Gamma \nabla^2 \delta(\vec r - 
                               \vec r') \delta(t-t')
\end{equation}
where $T$ is the temperature of the mixture, and $\langle ... \rangle$ 
denotes the
ensemble average.
The term $\vec \nabla \cdot (\varphi \vec v)$ describes the advection of 
the field
$\varphi$ by the velocity
and $\Gamma \nabla^2  (\delta {\cal F}/\delta \varphi)$ takes into account the 
diffusive transport of $\varphi$.
The free-energy functional generally used in phase separation studies 
of binary mixtures\cite{bray2} is 
\begin{equation}
\begin{cal} F \end{cal}= \int d {\vec r}\left[ \frac{a}{2}\varphi^2
+\frac{b}{4}\varphi^4+\frac{\kappa}{2} (\nabla \varphi)^{2}\right]
\label{fren}
\end{equation}
The first two terms are related to the bulk
properties of the fluid.  
While the parameter $b$ is always positive,
the sign of $a$ distinguishes between a disordered 
and a segregated mixture with phase-separated components.
The case with $a > 0$ gives a polynomial with one minimum in the origin
corresponding to the a disordered state with $\varphi=0$ everywhere
while two symmetric minima are present when $a < 0$ corresponding to the two pure phases
with $\varphi = \pm \sqrt{-a/b}$. These are the equilibrium values of the order parameter
$\varphi$ in the small temperature limit.
In the following we
 will consider a deep quench at $T=0$
into the two-phase coexistence region. 
The qualitative behaviour in the non-zero temperature case is similar: Thermal 
fluctuations are responsible for the increased roughness of the interfaces with respect
to $T=0$ and for the thermal excitations inside the ordered domains \cite{corberi2}.
The gradient term is related to the interfacial properties and takes into
account the energy cost for interface formation.
The  equilibrium profile between the two coexisting bulk phases
is (in a one-dimensional system) 
$\displaystyle \varphi(x)=\tanh \sqrt{\frac{1}{2 \kappa}}x$, giving a
 surface tension equal to   $\displaystyle \frac{2}{3} \sqrt{2 \kappa}$
and  an interfacial width
 proportional to $\sqrt{2 \kappa}$\cite{rowlinson}. 

The thermodynamic properties of the fluid 
follow from the free energy (\ref{fren}).
The chemical potential difference between the two fluids is given by
\begin{equation}
\frac{\delta{\cal F}}{\delta\varphi}=
a \varphi + b \varphi^3 - \kappa \nabla^2 \varphi .
\label{mu}
\end{equation}   
Equation (\ref{eqn2}) can be cast
in a dimensionless form after a redefinition of time, 
space and  field 
scales\cite{par}. Then we have chosen $\Gamma=|a|=b=\kappa=1$. 

The main observable for the study of the growth kinetics is 
the structure factor 
$C(\vec k,t)= \langle \psi(\vec k, t)\psi(-\vec k, t)\rangle $,
namely the Fourier transform of the real-space equal time correlation
function. 
$\psi(\vec k, t)$ is the Fourier transform of the concentration fluctuations
with respect to its average value: $\varphi - \langle \varphi \rangle$.
From the knowledge of the structure factor 
one computes the average size 
of domains. 
In the unsheared case ($\gamma = 0$ in Eq. (\ref{eqn2}))
this quantity can be defined as
\begin{equation}
R(t) = \pi \frac{ \int d k \;\;C(k,t)}{\int d k \;\;k  
\;\;C(k,t)}
\label{eqnradno}
\end{equation}
where 
\begin{equation}
C(k,t) = \frac{1}{N_k}\sum_{|\vec k|=k} C(\vec k,t)
\label{eqnc}
\end{equation}
is the spherical average of the structure factor and $N_k$ is the number of 
lattice vectors of length $k$ in the reciprocal space of momenta.
In the case with shear, due to anisotropy, we measure
\begin{equation}
R_x(t) = \pi \frac{ \int d\vec k C(\vec k,t)}{\int d\vec k |k_x|  
C(\vec k,t)}
\label{eqnrad}
\end{equation}
and analogously for the other directions. 
An excess of viscosity $\Delta \eta$ can be measured in sheared mixtures due to the work spent in
stretching the domains. $\Delta \eta$, which is of experimental interest,
is defined as\cite{onuki}
\begin{equation}
\Delta \eta = -\frac{1}{\gamma}\int 
\frac {d\vec k}{(2\pi)^d} k_x k_y C(\vec k,t) .
\label{eqnex}
\end{equation}
where $d$ denotes the number of spatial dimensions.

We have simulated Eq. (\ref{eqn2}) in $d=2$ by a first-order
Euler discretization scheme. 
Periodic boundary conditions have been implemented
in the $x$ direction; Lees-Edwards 
boundary conditions\cite{lees} were used in the $y$ direction (shear 
direction). 
These  boundary conditions, originally developed for molecular dynamic 
simulation of fluids in shear, 
consist of moving the top and bottom periodic images of the lattice
with respect to it so that they
require the identification of the 
point at $(x,0)$ with the one located at $(x+\gamma L \Delta t,L)$,
where $L$ is the size of the lattice (the same in all the directions) and
$\Delta t$ is the time discretization interval. 

\section{Results} \label{results}

Simulations were run using lattices of size 
$L=1024$ with grid size $\Delta x=0.5, 1$. 
We do not observe significative differences between these two choices
of $\Delta x$. The results here
shown were obtained with $\Delta x=1$, $\Delta t=0.01$,
$T=0$ and three different values of
$\gamma = 0, 0.00488, 0.0488$. The system was initialized
in a high temperature disordered state 
with $\varphi(\vec r,0)= \varphi_0 + f(\vec r)$, where $f(\vec r)$ is a random
number in the range $(-0.01, 0.01)$ such that 
$\langle \varphi(\vec r,0) \rangle = \varphi_0$.
The evolution was studied with
$\varphi_0=0.4$, corresponding to a composition
$70:30$ of the mixture.

\subsection{Unsheared quench}

In this section we report the results for the unsheared quench. 
In Fig. 1 we show a sequence of configurations at consecutive times. 
The minority phase is made of isolated 
droplets whose number decreases, while their average radius increases in time.
The main physical mechanism for coarsening in
the present case, when hydrodynamics is neglected and thermal fluctuations
are absent, is the evaporation-condensation mechanism
proposed by  Lifshitz and Slyozov\cite{lifshitz}. 
In this case
larger domains grow at the expenses of smaller domains due to diffusion of
atoms. This gives a growth  exponent $\alpha=1/3$. To check this prediction we 
computed the average size of domains $R(t)$ using Eq. (\ref{eqnradno}).
The plot showing the time dependence of $R(t)$ is presented in Fig. 2. After
an initial regime when well defined domains are forming, 
corresponding to the fact that $\varphi$ is approaching 
its equilibrium values either $1$ or $-1$, 
$R(t)$ shows from $t \sim 500$ onwards an asymptotic
growth consistent with an exponent $\alpha=1/3$. 
Due to the large system used we could eliminate
finite-size effects and observe the $1/3$ regime 
for two decades in time having a convincing evidence of the mentioned
growth mechanism.

\subsection{Sheared quench}

The effect of the advective term, due to shear, on the segregation process
is investigated in this section. We used two values of the shear rate 
$\gamma = 0.00488, 0.0488$. In Figs. 3-4 a time sequence of configurations for
each value of $\gamma$ is shown. Snapshots are at the same values of the 
shear strain $\gamma t$ in order to have an immediate comparison between
the two runs. 
For strains less than or of the order of one the effects of shear on the 
shape of bubbles are negligible: the mean sizes of the domains for a given 
shear rate remain directionally independent.  
However for the smaller shear rate
the system generates bigger droplets because there is more time, keeping
the strain fixed, for growth
before the shear can influence the morphology.
 
This  behaviour can be observed quantitatively computing the typical sizes
of domains according to Eq. (\ref{eqnrad}). 
The evolution of $R_x, R_y$ vs. $\gamma t$ is shown in Figs. 5-6.
The ratio $R_x/R_y$ stays constant ($\simeq 1$) until $\gamma t \leq 1$
for both the shear rates, then a crossover is observed. Moreover, we see that
at $\gamma t \simeq 1$, $R_y(\gamma=0.00488) \simeq 3.7$ and 
$R_y(\gamma=0.0488) \simeq 2.6$.

When the strain is larger than one,
there is a competition between coarsening and shear-induced deformations of
domains. For the smaller shear rate case, the diffusion-evaporation mechanism
is still important and only
domains which could grow in size are affected by the flow,
while smaller ones evaporate. 
When the shear rate is larger, on the other hand, 
patterns grow smaller compared
to those in the smaller shear rate case and 
diffusion from small to large droplets is inhibited as it can be seen in
Fig. 4 at $\gamma t = 3$, 
where the minority phase is more 
dispersed, consisting of a large number of domains with smaller surfaces.
In both the cases the growth is faster in
the flow direction and domains assume a striplike shape aligned with the flow.
As the elongation increases, non-uniform patterns appear. This phenomenon
was observed also in the case of sheared critical quenches\cite{corberi2} and
was interpreted as due to the presence in the system of two typical length
scales in each direction, which compete in time. At $\gamma t = 8$ there is
an abundance of stretched domains which is also reflected in the local 
minima of $R_y$ (see Figs. 5-6). A convincing evidence of such a behaviour
is given by the structure factor $C(\vec k,t)$ which is contour-plotted
at $\gamma t = 8$ in Fig. 7 for the case $\gamma=0.00488$ (a similar
behaviour is observed also for the other value of $\gamma$). It is separated
in two distinct foils, each of them having two peaks. 
The presence of a four-peaked structure factor in sheared mixtures 
is a quite general feature which can be also shown mathematically 
\cite{corberi1}.
Due to the
symmetry $C(\vec k,t)=C(-\vec k,t)$, the foils are symmetric and one can
consider only the two peaks of one foil. 
Let us consider the peaks which are in the
half-plane $k_x<0$. 
The peaks are located at $(k_{x_1}, k_{y_1}) \simeq (-0.035, -0.146)$ 
and $(k_{x_2}, k_{y_2}) \simeq (-0.080,0.310)$. 
The interpretation of a peak
in the structure factor as the signature of the existence in the system
of a characteristic length scale, indicates that there are two relevant
length scales for each direction in the present case proportional to 
$\pi/|k_{x_{1,2}}|$ and $\pi/|k_{y_{1,2}}|$ in the flow and in the shear direction,
respectively. 
This gives a value of around 2.3 and 2.1 for the ratio
between the characteristic sizes of domains in the flow and in the shear 
direction, respectively. In the case of
sheared quench of a symmetric mixture these ratios were 2.5 and
2.1 \cite{corberi2}. 
Considering the 
behaviour in the shear direction, we see that there are domains with two 
different thicknesses in the system (the behaviour in the flow direction is analogous).
At $\gamma t=8$ the peak which prevails
is the
one located at $(k_{x_2}, k_{y_2})$,
clearly indicating a larger abundance of thin domains, as already observed.
Then, when the strain increases, domains are stretched and a cascade
of ruptures occurs in those regions where the stress is higher and thicker
domains, not yet broken, prevail as at $\gamma t=15$.
This mechanism produces an oscillation in time in the behaviour of the peaks
of the structure factor and now it is the other peak, located at $(k_{x_1}, k_{y_1})$, to prevail. 
This is  shown in Fig. 8 where a three-dimensional plot of one foil of the structure factor is pictured 
at two consecutive times. 
This mechanism is similar to the one observed in the case of symmetric mixtures under shear flow
\cite{corberi2}.
Such an oscillation affects also $R_y$
whose amplitude is reduced at higher shear rates. We investigated also
the existence of a scaling regime, analogously to the one observed in the 
unsheared case. Due to the inevitable finite-size effects, simulations cannot 
give  a convincing evidence of such a scaling regime. We showed in a previous 
paper\cite{corberi2}, using a renormalization group scheme, that if there is
scaling, then $R_x \sim \gamma t^{4/3}$ and $R_y \sim t^{1/3}$. This means
that the growth in the shear direction is not affected by the external
velocity while
in the flow direction there is an extra contribution of $1$ to the growth
exponent coming from the advective term in Eq. (\ref{eqn2}). Our runs do not
allow to check this prediction. 

We also considered the rheological properties of the mixture. 
The excess viscosity 
was measured according to Eq. (\ref{eqnex}). 
The plots of $\Delta \eta$ as a function of $\gamma t$ are shown in
Figs. 9-10.
Starting from zero, $\Delta \eta$
shows a net increase up to a global maximum at $\gamma t \simeq 5$, 
independent of the shear rate. This results from the stretching of domains
caused by shear, which requires work against the surface tension. In the case
with $\gamma=0.00488$ (Fig. 9) this increase is not monotonic 
and $\Delta \eta$ shows
a first peak at $\gamma t < 1$.  This first peak disappears by increasing 
$\gamma$. The height of the main peak depends on the shear rate being
higher at smaller shear rates. This is due to the 
morphology of domains which have more interfaces for smaller $\gamma$
(cfr. Figs. 3-4 at $\gamma t = 5$), giving
a larger contribution to $\Delta \eta$. 
For larger times $\Delta \eta$ 
decreases due to the breakup of domains which dissipates energy previously 
stored. An oscillation is completed at $\gamma t \simeq 15$, which is
more evident for $\gamma=0.00488$, and it is related to the corresponding
oscillation in $R_y$. If scaling were obeyed, one would expect $\Delta \eta$
to scale as $\Delta \eta \sim R_x^{-1} R_y^{-1}$\cite{corberi2}. Given the
above discussed behaviour of $R_x$ and $R_y$, it should be $\Delta \eta
\sim t^{-5/3}$. This power law is roughly consistent with our data, especially
for $\gamma=0.00488$.

\section{Conclusions} \label{conclusions}

We studied the effects of a shear flow on the phase separation of  
binary mixtures with asymmetric composition. 
The binary mixture is described by a continuum free-energy functional
and we have solved numerically the related model B. 
Simulations were carried out on two-dimensional systems;  
lattices are large enough to minimize finite-size effects.

Without shear we observe an asymptotic scaling regime with a growth
exponent $\alpha=1/3$ consistent with previous theoretical\cite{yao} and 
numerical\cite{yeomans} studies. Thanks to the size
of the lattice, we could observe such an exponent over two time decades.

More interesting and rich is the phenomenology when an external flow is
applied during the quench. The patterns are stretched and aligned along the 
flow direction. Besides this behaviour, domain growth competes with 
shear-induced deformations and at high shear rates the minority phase
appears more dispersed and made of smaller domains. In the limit of very high
shear rates one could observe a complete demixing. Moreover, domains
with two different thickness are clearly visible in the simulations: we observe
a larger abundance first of thin and, then, of thicker domains, analogously
to what observed in sheared critical quenches\cite{corberi2}.  
The existence of two typical length scales in each spatial direction
is related to a four-peaked  structure factor.
The alternate predominance of these length scales produces
an oscillation between the peaks of the structure factor
 and in the physical observables of the system. 

The rheological properties were investigated through the excess viscosity.
This quantity increases up to a maximum and then decreases. This behaviour is
associated to the stretching of domains, 
which requires work against the surface
tension, and to the following ruptures of patterns which dissipates the stored
energy. The position of the main peak as a function of the strain
does not depend on the shear rate but
its height decreases increasing $\gamma$ due to the more dispersed morphology
of the minority phase.

Finally, we want to mention that we could not give a definitive answer
to the question, which is still matter of an open debate\cite{yeomans},
whether growth continues indefinitely with shear or a steady state is reached.
Related to this question there is the problem of a reliable determination
of the growth exponents.

{\bf Acknowledgments}

G.G. acknowledges partial
support by INFM PRA-HOP 1999. 
A.L. acknowledges partial support by INFM PA-G01-1.

\newpage

\begin{figure}
\centerline{\psfig{figure=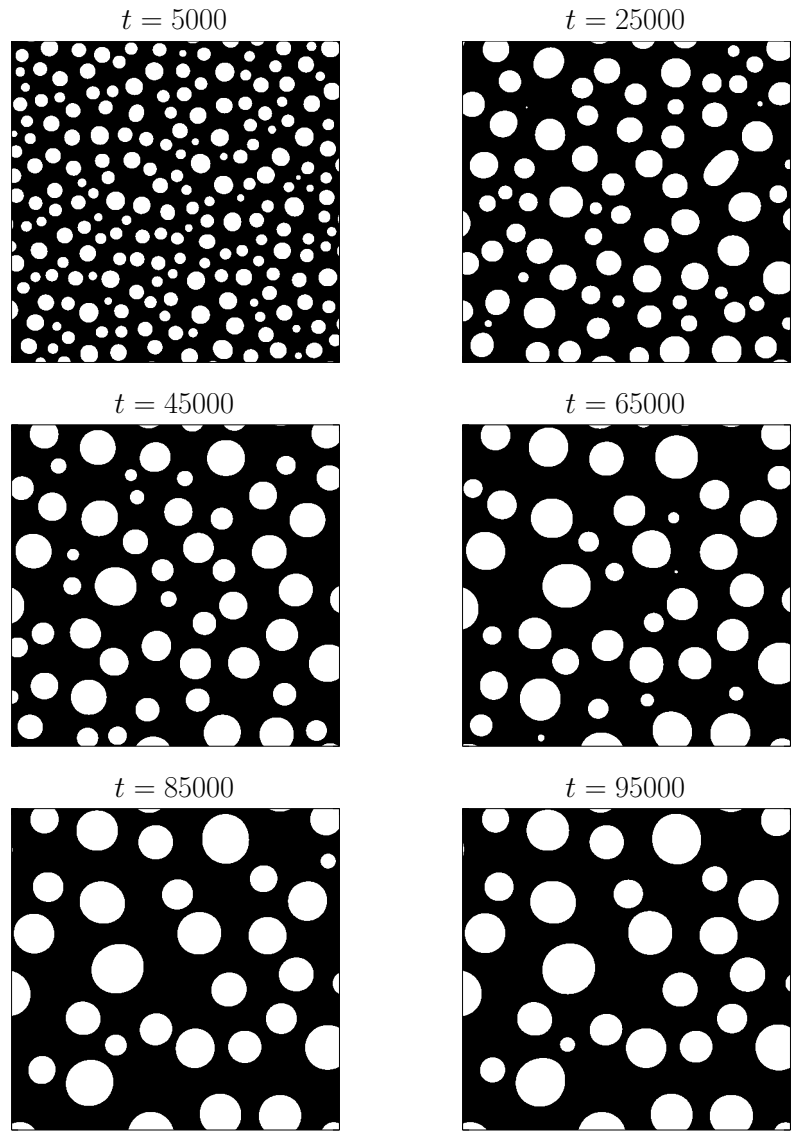,width=14cm,angle=0}}
\caption{Configurations of a portion of $512 \times 512$ sites of the
whole lattice are shown at different values of the time in the unsheared 
case.}
\label{confnoshear}
\end{figure}
\newpage

\begin{figure}
\centerline{\psfig{figure=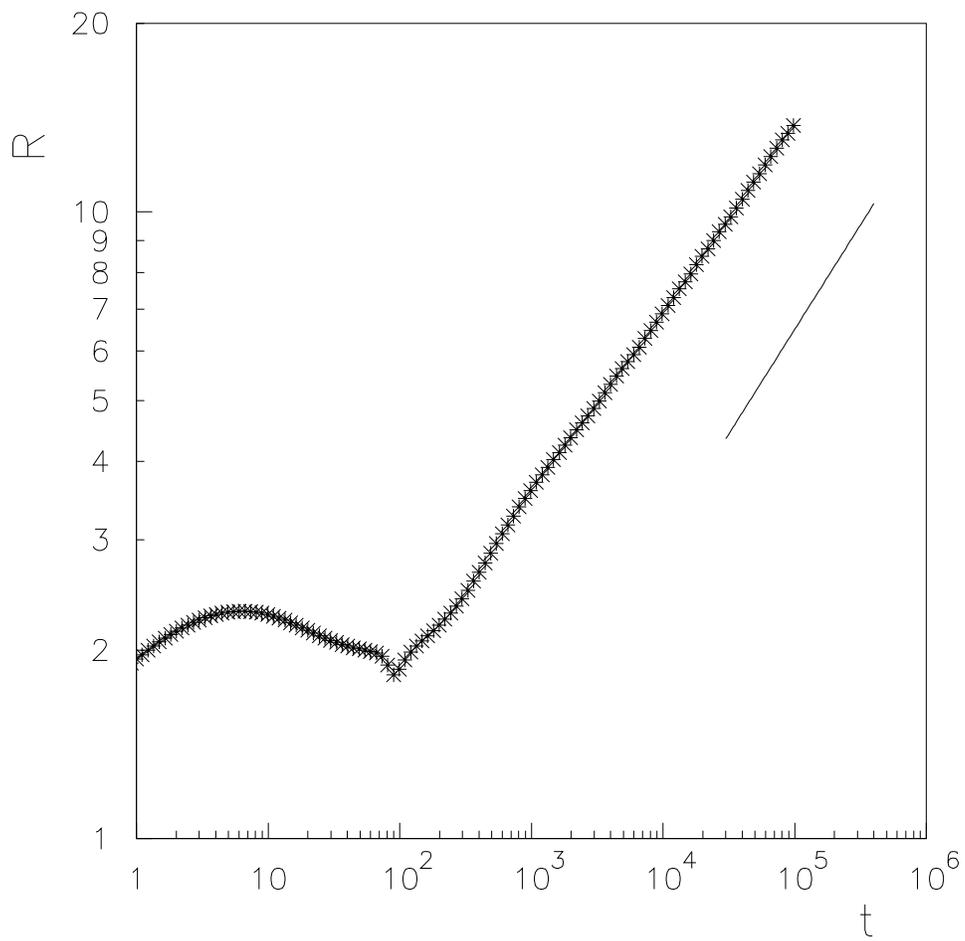,width=14cm,angle=0}}
\caption{Evolution of the average domain size in the unsheared case.
The straight line has slope 1/3.}
\label{radiusnoshear}
\end{figure}
\newpage

\begin{figure}
\centerline{\psfig{figure=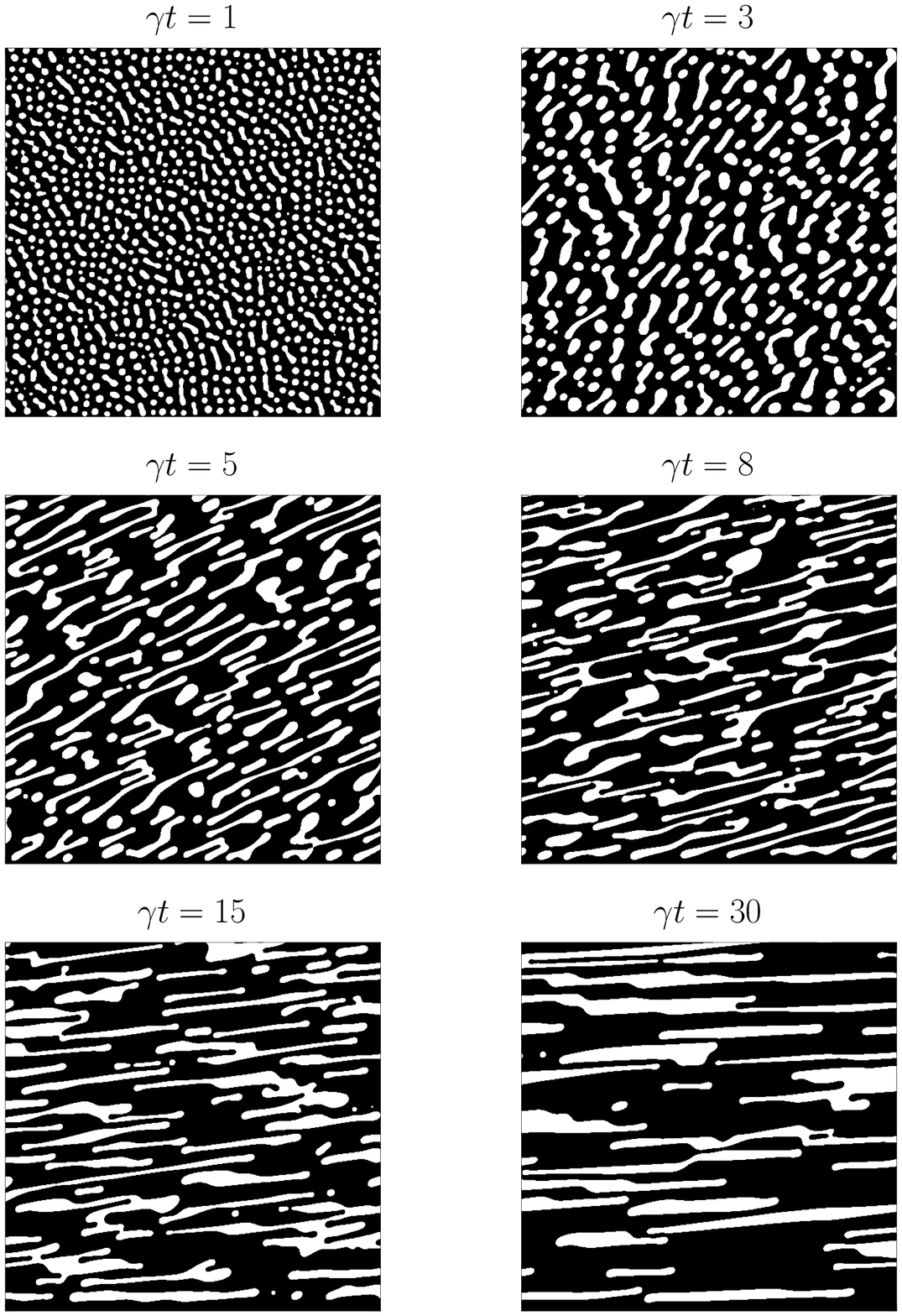,width=14cm,angle=0}}
\caption{Configurations of a portion of $512 \times 512$ sites of the
whole lattice are shown at different values of the shear strain
$\gamma t$ in the case with $\gamma = 0.00488$.
The $x$ axis is in the horizontal direction.}
\label{confshear1}
\end{figure}
\newpage

\begin{figure}
\centerline{\psfig{figure=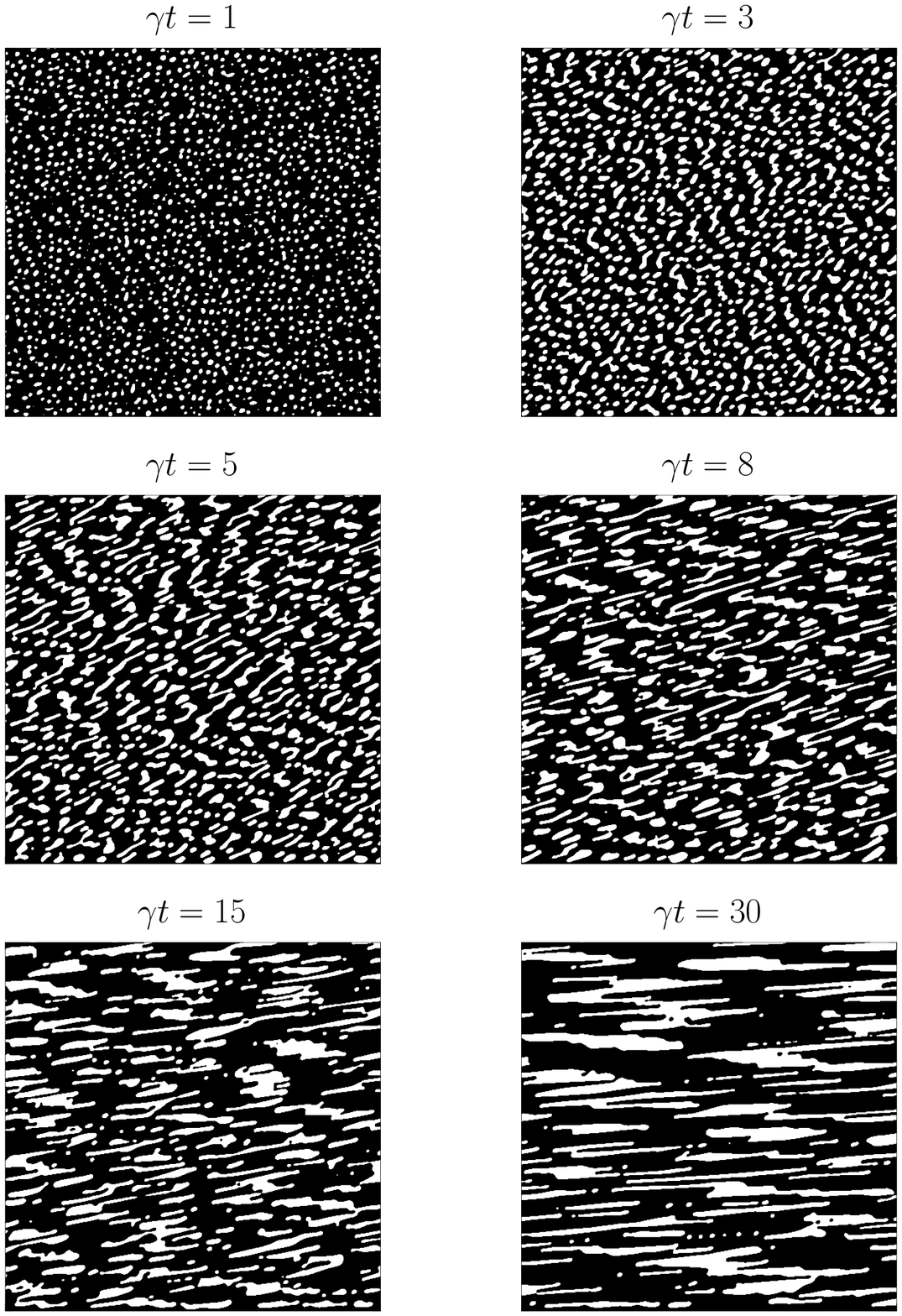,width=14cm,angle=0}}
\caption{Configurations of a portion of $512 \times 512$ sites of the
whole lattice are shown at different values of the shear strain
$\gamma t$ in the case with $\gamma = 0.0488$.
The $x$ axis is in the horizontal direction.}
\label{confshear2}
\end{figure}
\newpage

\begin{figure}
\centerline{\psfig{figure=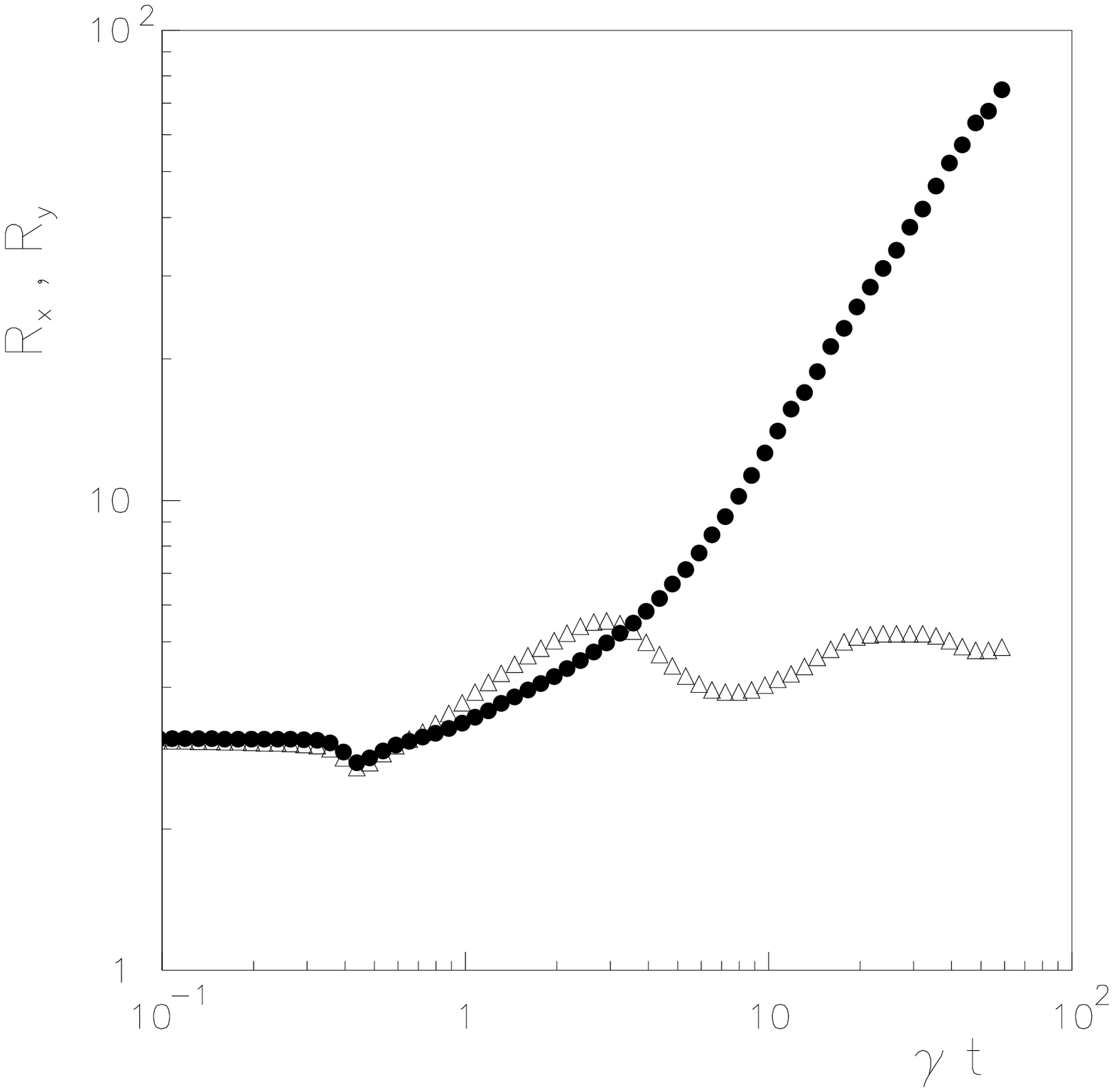,width=14cm,angle=0}}
\caption{Evolution of the average domains sizes in the shear (triangles)
and flow (circles) directions in the case with $\gamma = 0.00488$. 
The slope of $R_x$ is 1.1.}
\label{radiusshear1}
\end{figure}
\newpage

\begin{figure}
\centerline{\psfig{figure=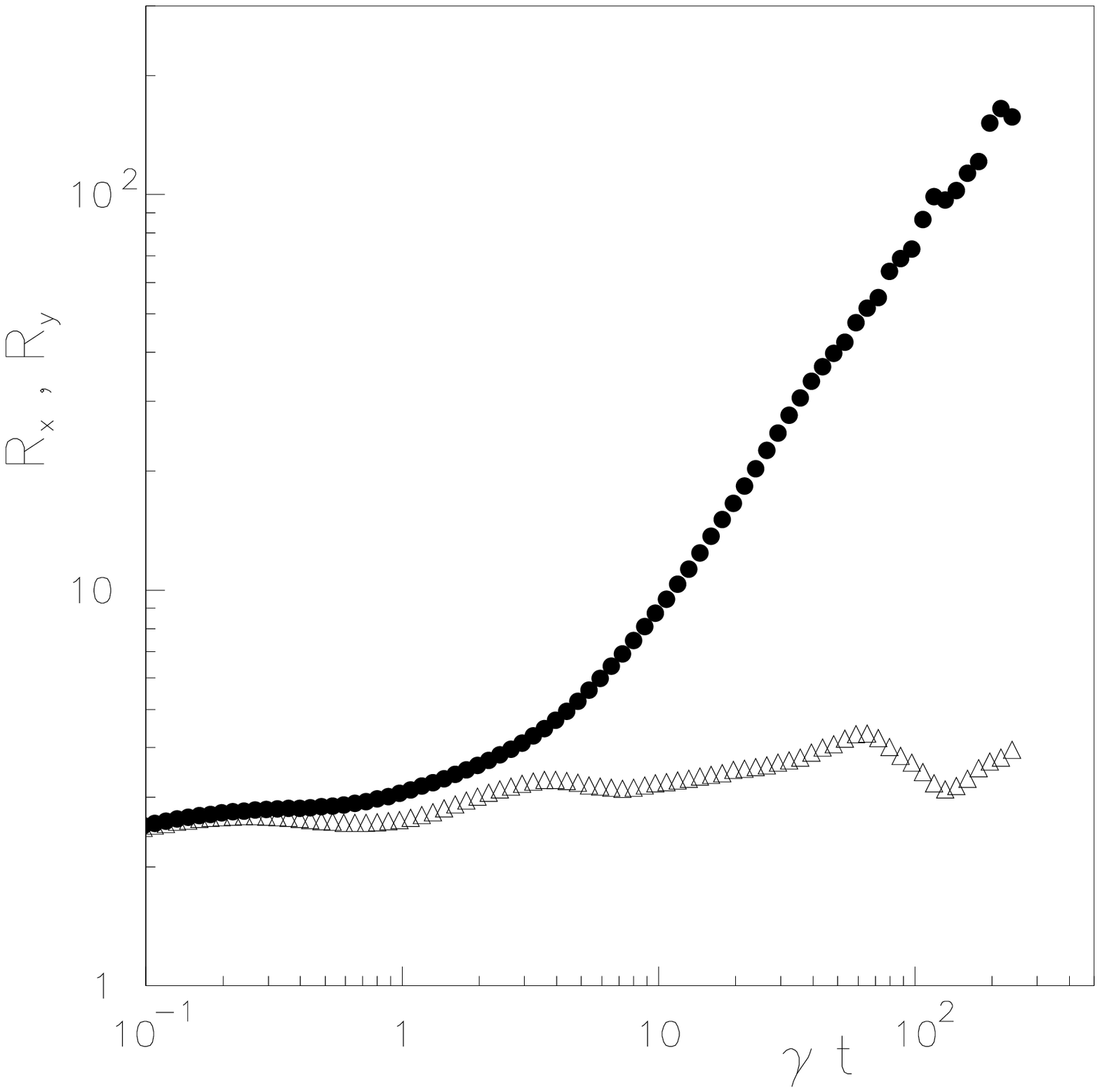,width=14cm,angle=0}}
\caption{Evolution of the average domains sizes in the shear (triangles)
and flow (circles) directions in the case with $\gamma = 0.0488$. 
The slope of $R_x$ is 1.1.}
\label{radiusshear2}
\end{figure}
\newpage

\begin{figure}
\centerline{\psfig{figure=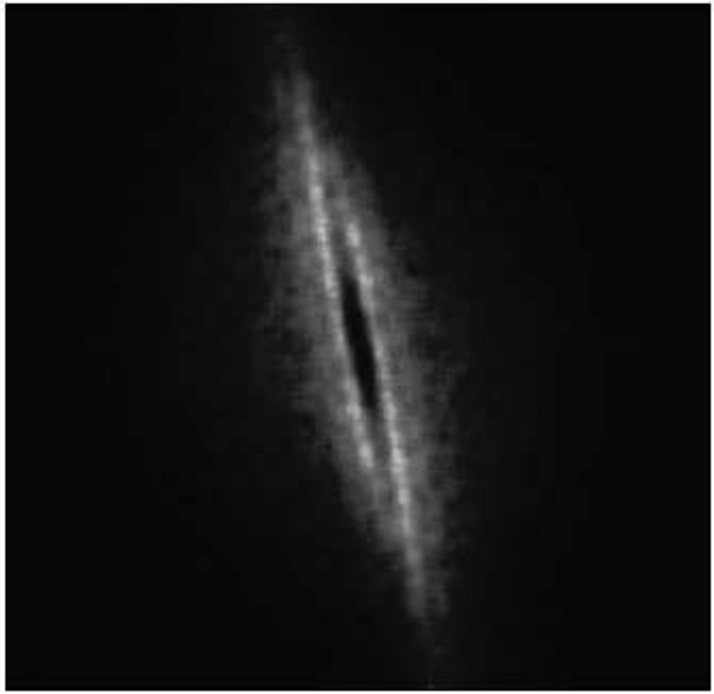,width=15cm,angle=0}}
\caption{Contour plot of the structure factor at $\gamma t = 8$
in the case with $\gamma = 0.00488$. 
The $k_x$ axis is in the horizontal direction, the $k_y$ axis is in the 
vertical direction.}
\label{fattstr1}
\end{figure}
\newpage

\begin{figure}
\centerline{\psfig{figure=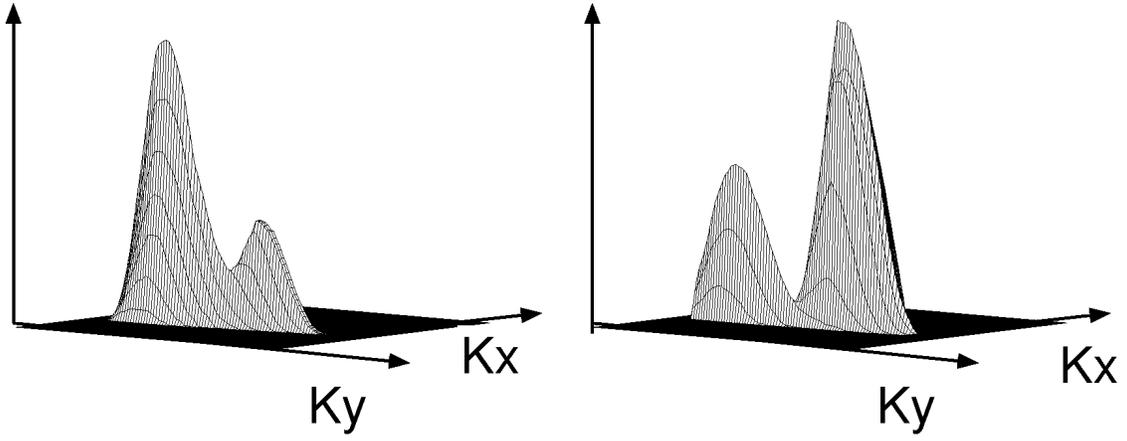,width=15cm,angle=0}}
\caption{Three-dimensional plot of the structure factor at $\gamma t = 8$ (left panel) and at
$\gamma t = 15$ (right panel)
in the case with $\gamma = 0.00488$. 
Only one foil of $C({\vec k},t)$ is shown (see the text for details).}
\label{fattbis}
\end{figure}
\newpage

\begin{figure}
\centerline{\psfig{figure=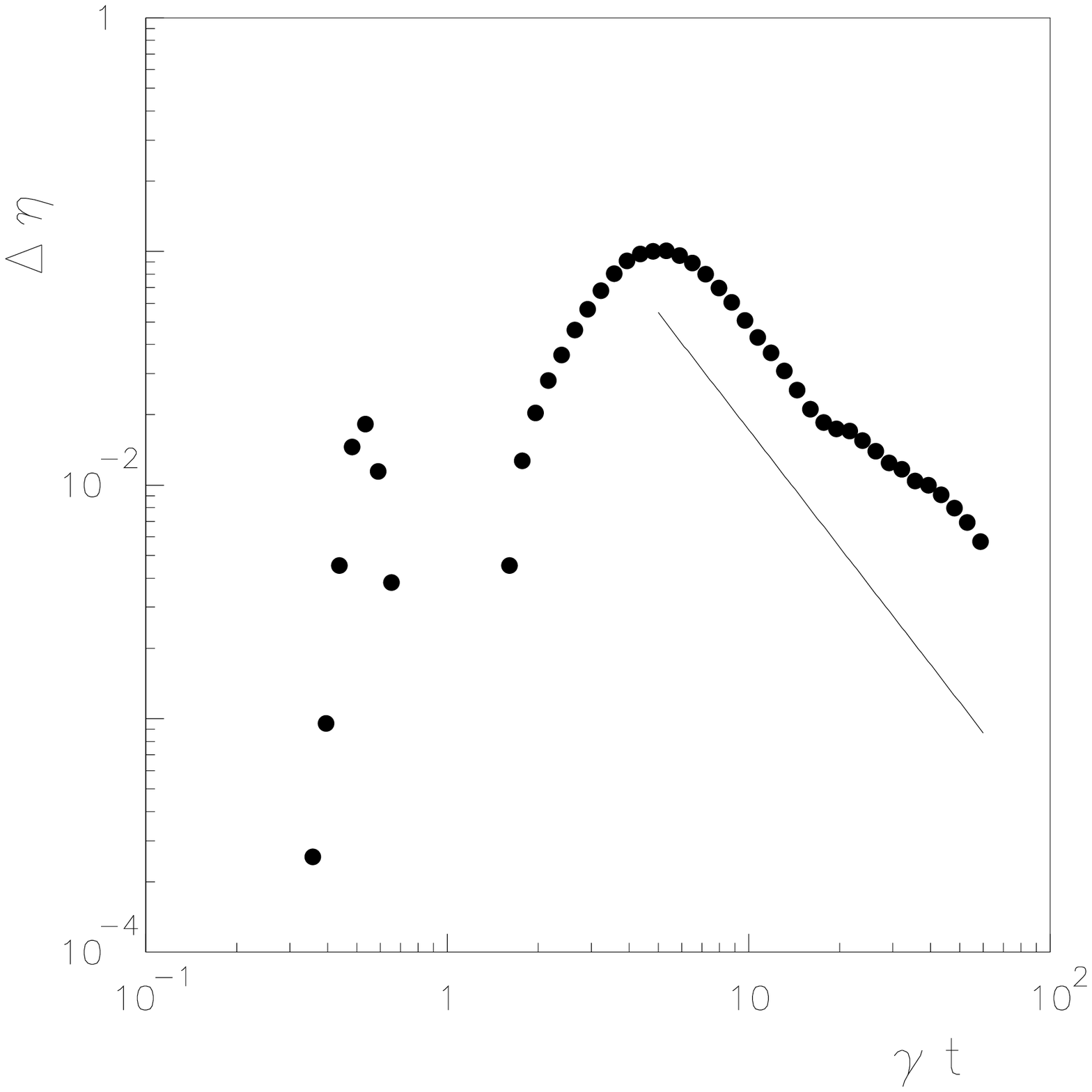,width=14cm,angle=0}}
\caption{The excess viscosity as a function of the shear strain in the 
case with $\gamma = 0.00488$. The slope of the straight line is -5/3.}
\label{exviscshear1}
\end{figure}
\newpage

\begin{figure}
\centerline{\psfig{figure=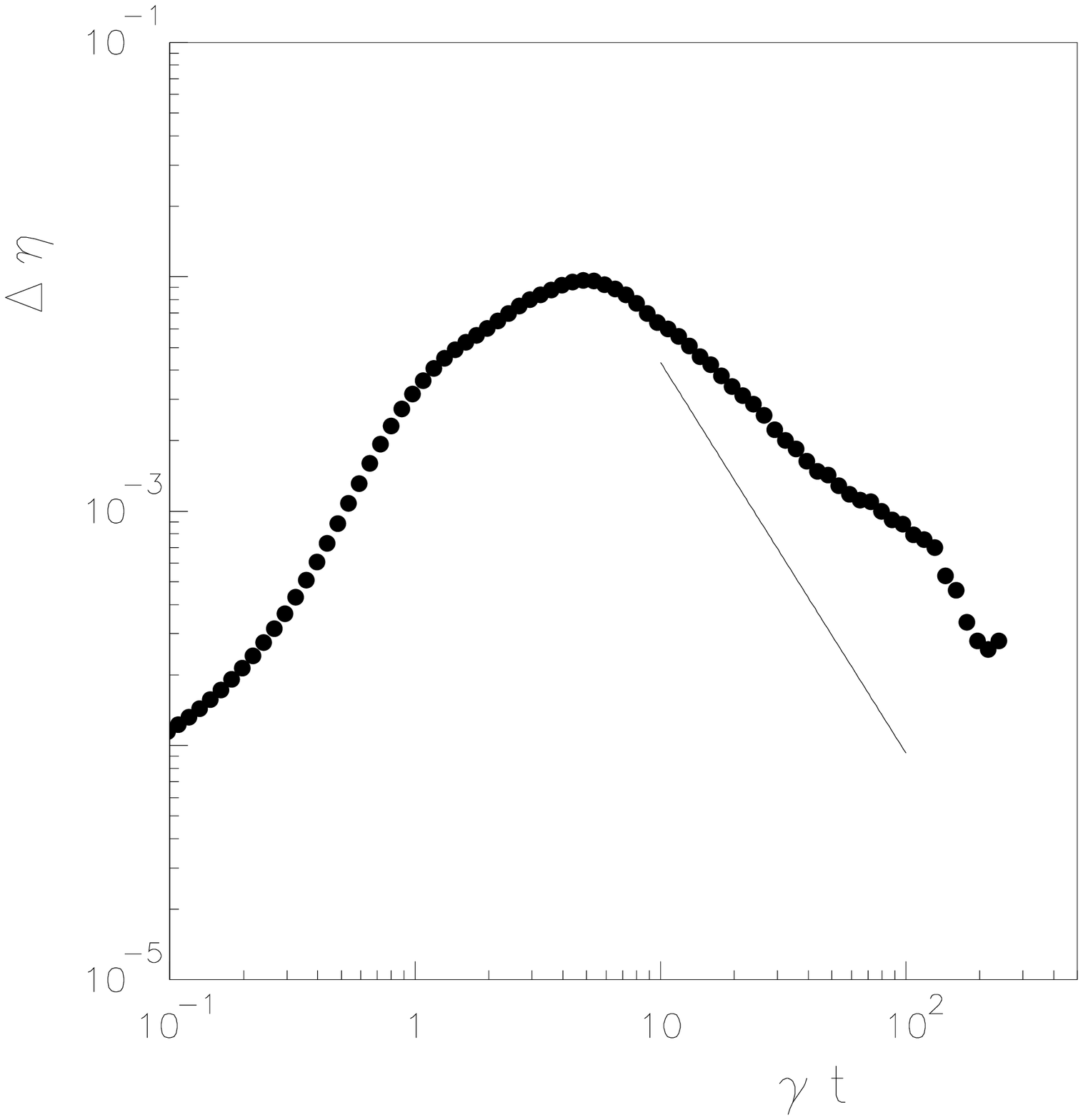,width=14cm,angle=0}}
\caption{The excess viscosity as a function of the shear strain in the 
case with $\gamma = 0.0488$. The slope of the straight line is -5/3.}
\label{exviscshear2}
\end{figure}

\end{document}